\documentclass[aps,prl,reprint,superscriptaddress]{revtex4-1}

\usepackage{amsmath,amsfonts,amssymb}

\usepackage[pdftex]{graphicx}
\usepackage{gensymb}
\usepackage{microtype}
\usepackage{bm}
\usepackage[normalem]{ulem}

\usepackage[svgnames]{xcolor}
\usepackage[
	colorlinks=True,linkcolor=DarkRed,citecolor=ForestGreen,urlcolor=MediumBlue,
	pdfstartview=FitH,bookmarks=False,pdfpagemode=UseNone
]{hyperref}

\definecolor{green}{rgb}{0.0, 0.7, 0.0}

\newcommand{\InSi}{In/Si(111)}
\newcommand{\Ga}{$\mathrm{\overline{\Gamma}}$}
\newcommand{\X}{$\mathrm{\overline{X}}$}
\newcommand{\Y}{$\mathrm{\overline{Y}}$}
\newcommand{\K}{$\mathrm{\overline{K}}$}
\newcommand{\EF}{$E_{\mathrm{F}}$}

\begin{document}

\title{\boldmath Excited state band mapping and momentum-resolved ultrafast population dynamics in In/Si(111) nanowires investigated with XUV based time- and angle-resolved photoemission spectroscopy}

\author{C. W. Nicholson}
\email{christopher.nicholson@unifr.ch}

\altaffiliation{Present address: Department of Phyiscs, University of Fribourg, Chemin du Mus\'ee 3, 1700 Fribourg, Switzerland}
\affiliation{Fritz-Haber-Institut der Max-Planck-Gesellschaft, Faradayweg 4-6, Berlin 14915, Germany}
\author{M. Puppin}
\altaffiliation{Present address: Laboratory of Ultrafast Spectroscopy, \'Ecole Polytechnique F\'ed\'erale de Lausanne, CH-1015 Lausanne, Switzerland}
\affiliation{Fritz-Haber-Institut der Max-Planck-Gesellschaft, Faradayweg 4-6, Berlin 14915, Germany}
\author{A. L\"ucke}
\affiliation{Unversity of Paderborn, Warburger Strasse 100, 33098 Paderborn, Germany}
\author{U. Gerstmann}
\affiliation{Unversity of Paderborn, Warburger Strasse 100, 33098 Paderborn, Germany}
\author{M. Krenz}
\affiliation{Unversity of Paderborn, Warburger Strasse 100, 33098 Paderborn, Germany}
\author{W. G. Schmidt}
\affiliation{Unversity of Paderborn, Warburger Strasse 100, 33098 Paderborn, Germany}
\author{L. Rettig}
\affiliation{Fritz-Haber-Institut der Max-Planck-Gesellschaft, Faradayweg 4-6, Berlin 14915, Germany}
\author{R. Ernstorfer}
\affiliation{Fritz-Haber-Institut der Max-Planck-Gesellschaft, Faradayweg 4-6, Berlin 14915, Germany}
\author{M. Wolf}
\email{wolf@fhi-berlin.mpg.de}
\affiliation{Fritz-Haber-Institut der Max-Planck-Gesellschaft, Faradayweg 4-6, Berlin 14915, Germany}

\date{\today}

\begin{abstract} 
We investigate the excited state electronic structure of the model phase transition system In/Si(111) using femtosecond time- and angle-resolved photoemission spectroscopy (trARPES). An extreme ultraviolet (XUV) 500~kHz laser source at 21.7~eV is utilized to map the energy of excited states above the Fermi level, and follow the momentum-resolved population dynamics on a femtosecond time scale. Excited state band mapping is used to characterize the normally unoccupied electronic structure above the Fermi level in both structural phases of indium nanowires on Si(111): the metallic (4x1) and the gapped (8x2) phases. The extracted band positions are compared with the band structure calculated within density functional theory (DFT) within both the LDA and GW approximations. While good overall agreement is found between the GW calculated band structure and experiment, deviations in specific momentum regions may indicate the importance of excitonic effects not accounted for at this level of approximation. To probe the dynamics of these excited states, their momentum-resolved transient population dynamics are extracted with trARPES. The transient intensities are compared to a simulated spectral function modeled by a state population employing a transient elevated electronic temperature as determined experimentally. This allows the momentum-resolved population dynamics to be quantitatively reproduced, revealing important insights into the transfer of energy from the electronic system to the lattice. In particular, a comparison between the magnitude and relaxation time of the transient electronic temperature observed by trARPES with those of the lattice as probed in previous ultrafast electron diffraction studies implies a highly non-thermal phonon distribution at the surface following photo-excitation. This suggests that the energy from the initially excited electronic system is initially transferred to high energy optical phonon modes followed by cooling and thermalization of the photo-excited system by much slower phonon-phonon coupling.

\end{abstract}

\maketitle

\section{I. Introduction}

The use of time-resolved spectroscopies to create and probe matter out-of-equilibrium offers the chance to study a range of fundamental dynamical processes including chemical reaction dynamics \cite{Zewail2000, Dell'Angela2013, Ostrom2015}, atomic motion at surfaces \cite{Petek2000, Vogelgesang2018} or the ultrafast flow of energy within materials \cite{Bovensiepen2007, Waldecker2016, Feist2018}. In complex solid state systems with coupled degrees of freedom, access to both energy ($E$) and momentum ($k$) is particularly useful for understanding the dynamic changes to the electronic structure that occur following photo-excitation. Angle-resolved photoemission spectroscopy (ARPES) \cite{Hufner1995, Damascelli2003, Hufner2007} is a powerful technique that gives access to the single-particle spectral function $A$($E$, $k$) but is restricted to states below, or on the order of $k_{B}T$ above, the Fermi level (\EF{}). Extending this technique to the femtosecond time scale with pump-probe trARPES allows simultaneous access to normally unoccupied states and occupied states below \EF{}. Mapping the energy dispersion of bands above \EF{} significantly extends the accessible energy range available with conventional ARPES, and as such enables an extended test of the applicability of current theoretical models in the prediction of material properties. 

\begin{figure}[ht]
\includegraphics[width=\columnwidth]{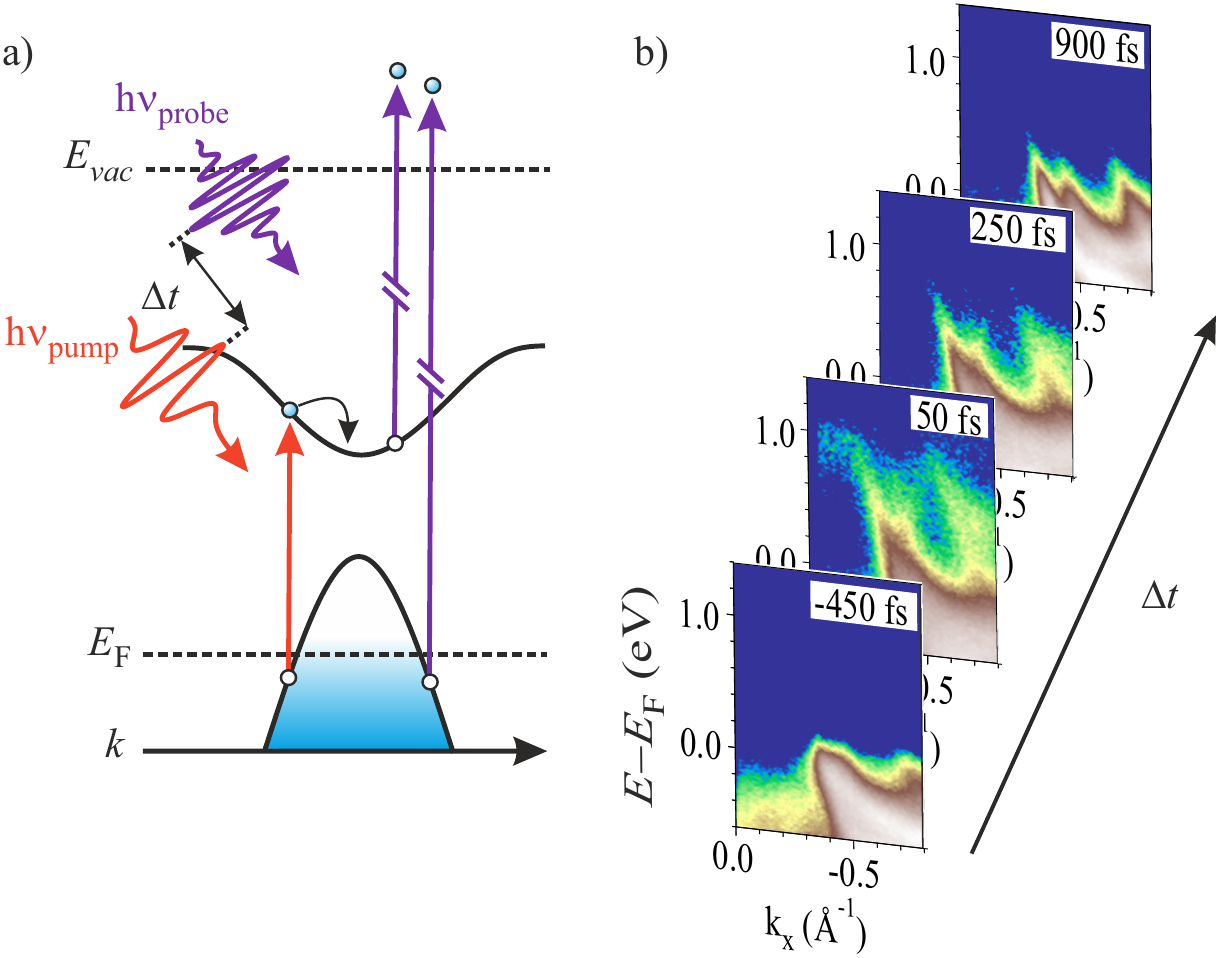}
\caption{\label{fig:trARPES} a) Schematic energy level diagram of the trARPES measurement scheme. Excitation by the pump pulse into the unoccupied states is followed by photo-removal by the probe pulse at variable delay. b) Time-resolved band structure of the In/Si(111) nanowires at different pump-probe delays. From such measurements both the dispersion in the unoccupied states and the energy and momentum-resolved population dynamics may be extracted.}
\end{figure}

An overview of the transient excited state data obtainable with trARPES is summarized in Fig.~\ref{fig:trARPES}. The schematic energy level diagram in a) shows how electrons from the occupied state region can be excited into the unoccupied states above \EF{} by the absorption of the pump photon. The probe pulse subsequently captures a snapshot of the transient distribution of the electrons at time $\Delta t$ after excitation by photo-exciting them above the vacuum level, after which they are detected by an electron analyzer. By fixing $\Delta t$ and varying the emission angle of the electrons with respect to the analyzer as in traditional ARPES measurements, the unoccupied electronic states can be mapped out in ($E, k_{x}, k_{y}$). These transiently populated states decay on a femtosecond time scale to lower lying energy levels by scattering with other electrons or other degrees of freedom e.g. phonons. By varying $\Delta t$, a full dynamic picture of the electronic state may be obtained, as presented in Fig.~\ref{fig:trARPES}~b). A fully dynamic spectral function $A(E, k, \Delta t)$ may therefore be obtained giving access to many-body coupling, ultrafast scattering pathways and lifetimes \cite{Kirchmann2010, Bovensiepen2012, Sobota2012, Yang2015}, as well as the possibility to follow band structure dynamics during photo-inducing phase transitions (PIPTs) \cite{Perfetti2006, Rini2008, Schmitt2008, Rohwer2011, Rettig2014, Monney2016}. 

A model phase transition system involving strongly coupled electron and phonon degrees of freedom are quasi-one dimensional indium nanowires formed at a silicon (111) surface: In/Si(111). These nanowires undergo a structural phase transition from a (4x1) to a (8x2) structure upon cooling below 130~K \cite{Yeom1999, Klasing2014}, concomitant with an insulator-to-metal transition \cite{Tanikawa2004, Sun2008}. Schematic diagrams of the real space structure in the two phases are shown in Fig.~\ref{fig:InSi}~a) and b). The high temperature metallic phase exhibits a characteristic ``zig-zag'' motif in real space, while in $k$-space there are three In bands with $p$-orbital character that cross \EF{}, labeled $m_{1}, m_{2}$ and $m_{3}$ in Fig.~\ref{fig:InSi}~c), located in the Si bulk band gap. The system transforms into a ``distorted hexagon'' with a gapped electronic structure in the low-temperature phase (Fig.~\ref{fig:InSi}~d)). The Peierls-like nature of the phase transition has long been debated \cite{Yeom1999, Yeom2002, Ahn2004, Lee2004, Gonzalez2006, Kim2016}. Theoretical investigations have suggested that the phase transition occurs through a combination of rotary and shear distortions \cite{Gonzalez2005, Riikonen2006, Wippermann2010, Jeckelmann2016}. The proposed Peierls mechanism \cite{Ahn2004} is that during the phase transition the initially metallic $m_{1}$ band at \X{} moves above \EF{}, transferring electrons to the $m_{2}$ and $m_{3}$, which become unstable towards the formation of a Peierls gap (Fig.~\ref{fig:InSi}~c). An observed softening of the relevant phonon modes \cite{Jeckelmann2016} further supported a Peierls-like mechanism driving the phase transition. However, our recent trARPES study, which charted the energy and momentum-resolved evolution of the band structure during the photo-induced phase transition \cite{Nicholson2018a}, suggests a behavior with several distinct time scales beyond a simple Peierls description of the phase transition. By photo-exciting the (8x2) phase, the insulator-to-metal transition in the $m_{2}$ and $m_{3}$ bands was observed to occur after 200~fs, while the $m_1$ band shifts from above to below the Fermi level within 500~fs. Moreover, the full structural PIPT was observed to occur after 700~fs, in agreement with previous ultrafast electron diffraction (UED) measurements \cite{Frigge2017}. An independent trARPES study carried out at 1~kHz repetition rate also reported a time-scale of 660~fs for the phase transition \cite{Chavez-Cervantes2018a}. A comparison with constrained DFT-based molecular dynamics simulations allowed us to extract the microscopic mechanism of the phase transition in terms of transient band populations. In particular, the key role played by zone boundary photo-holes was revealed, and enabled an description of the femtosecond dynamics of chemical bonds in real space. The importance of photo-holes for the PIPT was further supported by recent trARPES measurements employing sub-gap excitation \cite{Chavez-Cervantes2018c}. 

Here, we extend our previous investigations of the PIPT in In/Si(111) by carrying out a detailed analysis of the excited state band dispersions and transient population dynamics. Band mapping is carried out in both (4x1) and (8x2) phases and the resuts are compared with the expectations of DFT. We find that while results obtained within the GW approximation match very well with the majority of our measured spectra, there exist noticeable and important differences, which indicate that other factors such as excitonic effects, local defects or induced strain may be relevant in this reduced dimensional system. In addition, we investigate the momentum-resolved population dynamics within the excited state band structure, and compare to a simulated spectral function. By employing a transient quasi-thermal electronic temperature description, we are able to reproduce the markedly different populations evolutions as a function of energy and momentum, thus revealing the usefulness of such a simple ensemble description for out-of-equilibrium dynamics. Furthermore we show that such a verification of an electronically thermal model is important for understanding the energy flow from electrons to the lattice in the system. The fact that the highly excited electronic temperature is at odds with UED measurements, which derived a much smaller temperature increase of the surface In atoms of only $\sim$30~K for similar excitation conditions \cite{Frigge2018}, implies a highly non-thermal distribution of optical phonons at the surface, and a bottleneck for the cooling of the electronic system as a result of electron-phonon and phonon-phonon coupling.

\section{II. Experimental and Theoretical Methods}

\begin{figure}[b]
\includegraphics[width=\columnwidth]{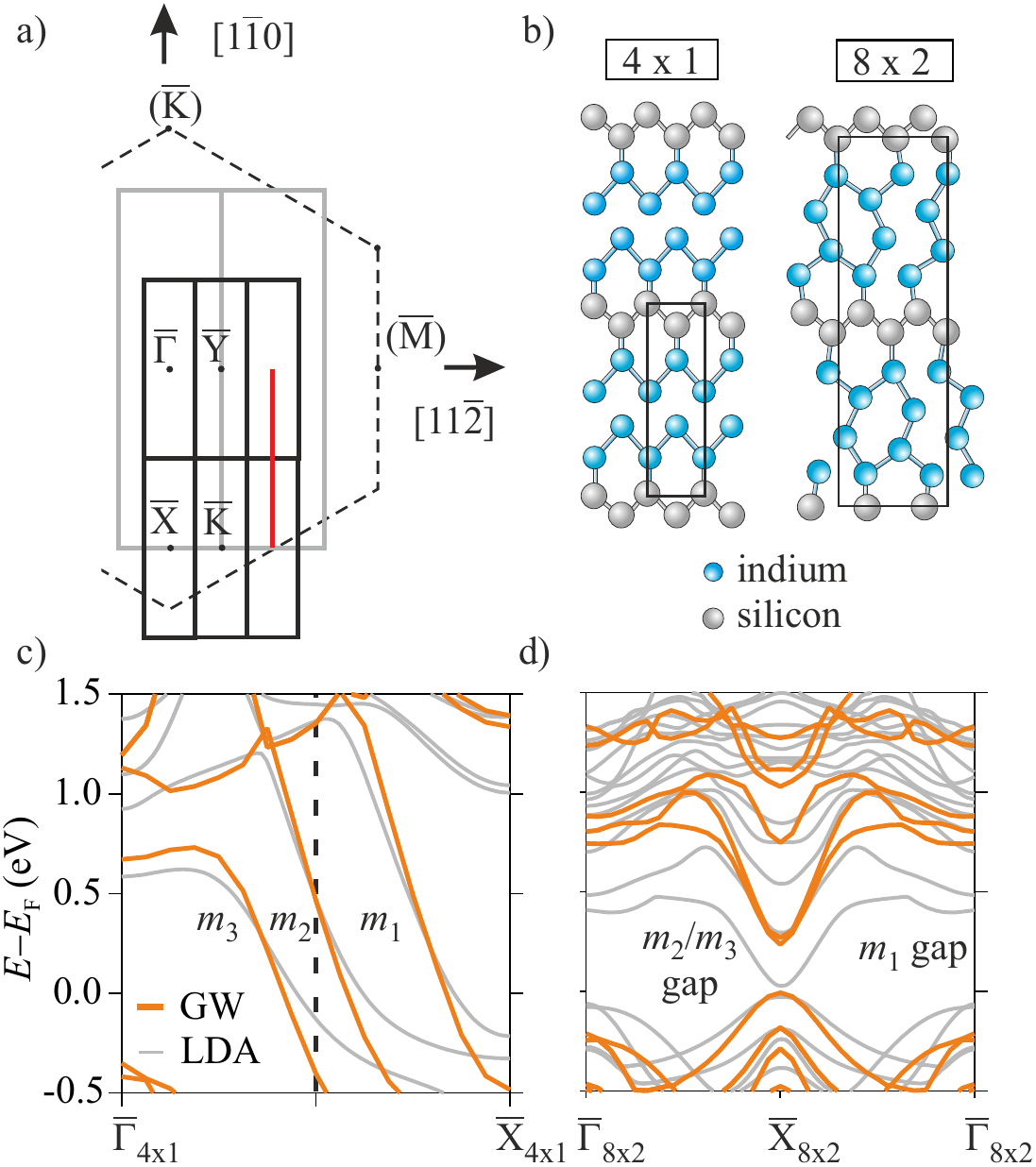}
\caption{\label{fig:InSi} (a) Brilloiun zone of the (4x1) (gray) and (8x2) (black) phases of In/Si(111). High-symmetry points are marked for the (4x1) phase. The dashed line marks the surface Brillouin zone of Si(111). The red solid line shows the \Ga{}-\X{} line along which the majority of data was obtained. (b) Schematic real-space structure after \cite{Wippermann2010} in the (4x1) phase and the (8x2) phase revealing the structural motifs of the two phases. The respective unit cells are marked. (c) Electronic band structure calculated within the GW (orange) and LDA (gray) approximations in the (4x1) phase and (d) in the (8x2) phase.
}
\end{figure}

\begin{figure}[b]
\includegraphics[width=\columnwidth]{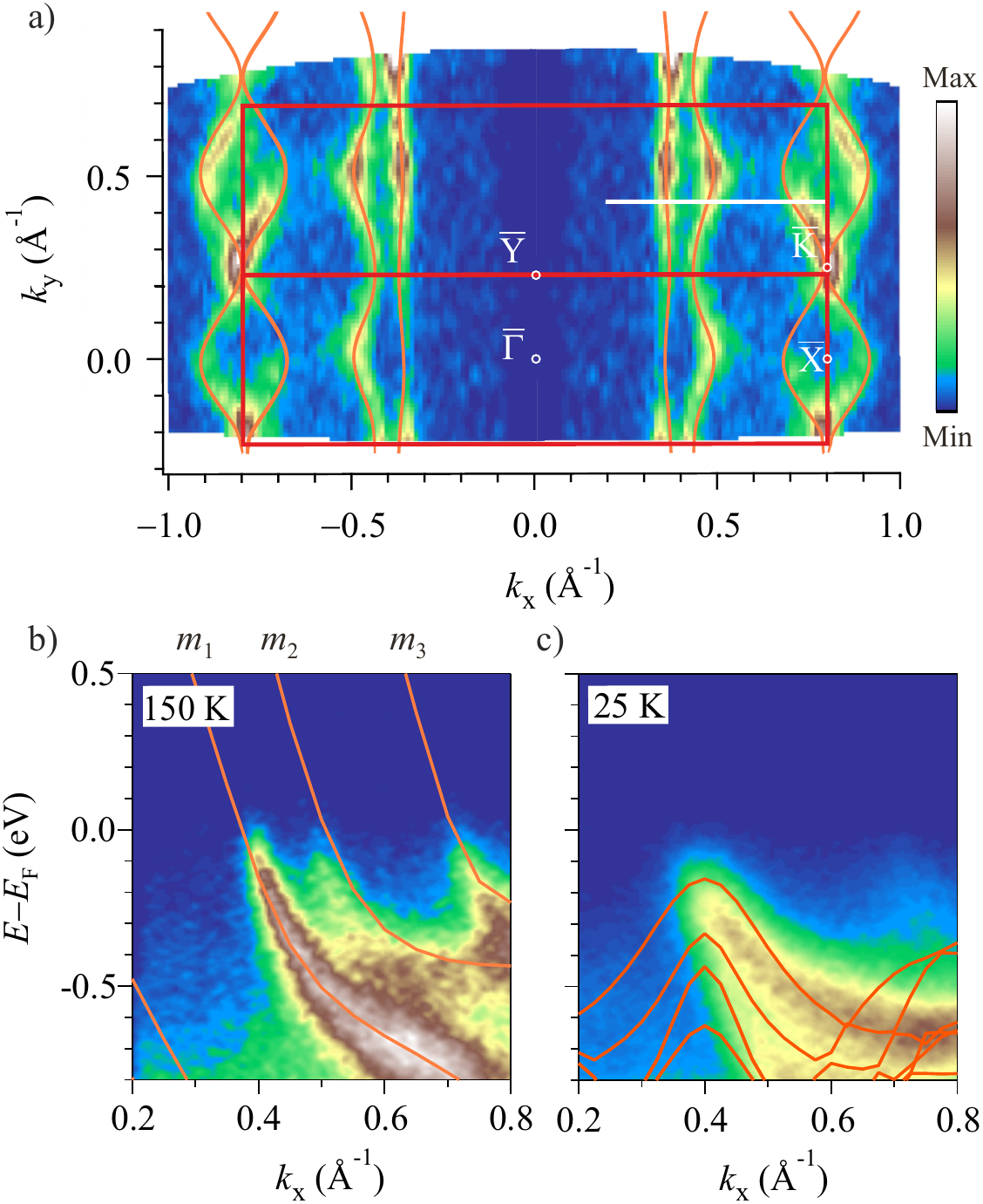}
\caption{\label{fig:thermalTrans} (a) Symmetrized Fermi surface obtained with the 21.7~eV HHG laser source at 150~K overlaid with the DFT Fermi surface sheets of the (4x1) phase (orange). The (4x1) Brillouin zone is overlaid (red). These results are consistent with previous studies \cite{Yeom1999,Ahn2004,Sun2008}. Data reproduced from Ref. \cite{Nicholson2018a}. (b) $E$ vs. $k_x$ cuts through the Fermi surface obtained at $k_{y}$=0.43\AA$^{-1}$ at 150~K revealing the three characteristic bands of the (4x1) phase. Calculated bands in the GW approximation along the \Ga$-$\X{} line are overlaid. (c) The same cut as in (b) at 25~K in the (8x2) phase revealing the gapped electronic structure and an asymmetric spectral weight. 
}
\end{figure}

In/Si(111) nanowires were grown epitaxially in ultra-high vacuum on a Si(111) substrate miscut by 2\degree{} towards the [-1 -1 2] direction. Substrates (MaTecK GmbH) were p-doped with boron to a resistivity of 0.075-0.085 $\Omega$ cm. Slightly more than 1 monolayer of pure In was evaporated from a home-built Knudsen cell at a rate of around 0.05 monolayers per minute onto a clean Si(111)-7x7 reconstructed surface at room temperature. Excess material was subsequently removed by direct current annealing at around 500 \degree C. The optimal coverage of 1 monolayer was judged live from the evolurion of the (4x1) pattern using low-energy electron diffraction (LEED).

In order to perform trARPES measurements across a wide $k$-space range with high statistics, we have developed a 500~kHz repetition rate XUV source at 21.7~eV, with a flux of 2x10$^{11}$ photons/s in a single harmonic \cite{Puppin2019} based on an optical parametric amplifier \cite{Puppin2015a}. The pump energy employed was 1.55~eV. A cross correlation between pump and probe of 40~fs and energy resolution of 150~meV were achieved in these experiments, both values obtained from the photoemission signal from the sample. The pump beam was linearly $s$-polarized and was incident at an angle of 15\degree{} to the sample surface normal. The pump and probe spot sizes were 300 x 170 $\mu$m$^2$ and 120 x 95 $\mu$m$^2$ respectively (FWHM). Given the high repetition rate of our laser system, we have analyzed average heating effects in our data. We find that for fluences below 3~mJ~cm$^{-2}$ such effects are less than 40~K, and therefore perform all measurements presented here in a fluence regime below this. ARPES and trARPES measurements are obtained with a 2D hemispherical analyzer (SPECS GmbH) in conjunction with a 6-axis cryogenic manipulator which can be cooled to 15~K (SPECS GmbH).

The ground-state atomic and electronic structure of In/Si(111) is determined with DFT within the local density approximation (LDA) using the Quantum Espresso implementation \cite{Giannozzi2009}. The surface is modeled using a supercell containing three bilayers of silicon, the bottom layer of which is saturated with hydrogen. The electron-ion interaction is modeled with norm-conserving pseudopotentials. Plane waves up to an energy cutoff of 50~Ry are used to expand the electronic orbitals. The surface Brillouin zone is sampled using a 2x8 Monkhorst-Pack mesh. Quasiparticle band structures for the (4x1) and (8x2) surface phases are obtained within the one-shot GW approximation \cite{Martin2016}, where the one-particle Green’s function, G, and the screened Coulomb interaction, W, are obtained from the LDA electronic structure. The band structures for the two phases presented in Fig.~\ref{fig:InSi}~c) and d) show overall similar structure, however the gap sizes at both the \Ga{} and \X-points are much smaller in the case of LDA. Such underestimation of electronic band gaps is typical of DFT calculations performed at the level of LDA.

The Fermi surface of In/Si(111) in the metallic (4x1) phase obtained with the XUV laser at 21.7~eV is presented in Fig.~\ref{fig:thermalTrans}~a). The data are symmetrized about the $k_{x}=$0 line in order to reveal band positions within the full unit cell which is shown along with the high-symmetry points. The quasi-one dimensional nature of the system is evident from the warped Fermi surface sheets, which imply a certain degree of inter-wire coupling \cite{Nicholson2017}. The measured data shows excellent agreement with previous studies \cite{Yeom1999, Ahn2004, Sun2008} and with the LDA calculated Fermi surface. We have characterized the differences between the two thermally stabilized phases, which are shown in Fig.~\ref{fig:thermalTrans}~b) and c) for the high and low-temperature phases respectively. At 150~K the characteristic bands $m_{1,2,3}$ of the (4x1) phase are clearly seen to disperse up to \EF{} in excellent agreement with our GW calculations. Upon cooling into the (8x2) phase, spectral weight is removed from \EF{} as the system becomes gapped. In the $m_{2}/m_{3}$ region, the resulting band dispersion turns away from \EF{}, with only weak spectral weight in the renormalized dispersion, as expected for charge density wave systems \cite{Voit2000}. The $m_{1}$ region also shows a significant decrease of intensity at \EF{}, although it is worth noting that a small amount of the original $m_{1}$ band intensity still persists, even at lowest temperatures, in contrast to the prediction of theory. Such signatures suggest the presence of small domains of (4x1) even at low-temperatures \cite{Lee2002, Lee2005, Guo2005, Morikawa2010}, which may be pinned at defects or step edges \cite{Wall2012}.

\begin{figure}[ht]
\includegraphics[width=\columnwidth]{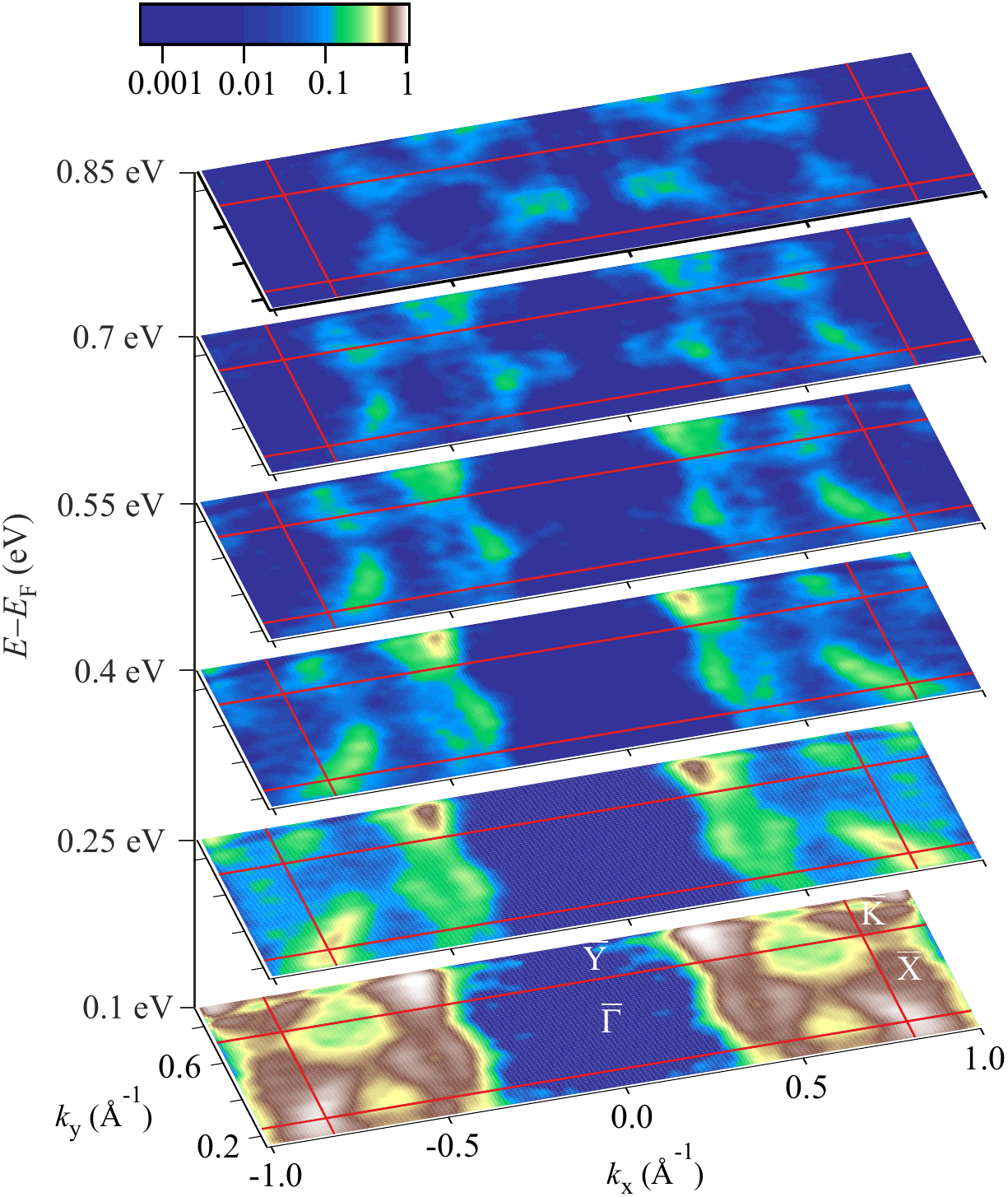}
\caption{\label{fig:mapping} An overview of the unoccupied band mapping in the (4x1) phase at 150~K obtained for a delay of 50~fs. The data are shown on a logarithmic color scale and have been symmetrized along the \Ga{}--\Y{} direction to show the full Brillouin zone.
}
\end{figure}

\section{III. Results and Discussion}

\subsection{1. Excited state mapping} \label{ssec:mapping}

\begin{figure*}[ht]
\includegraphics[width=2\columnwidth]{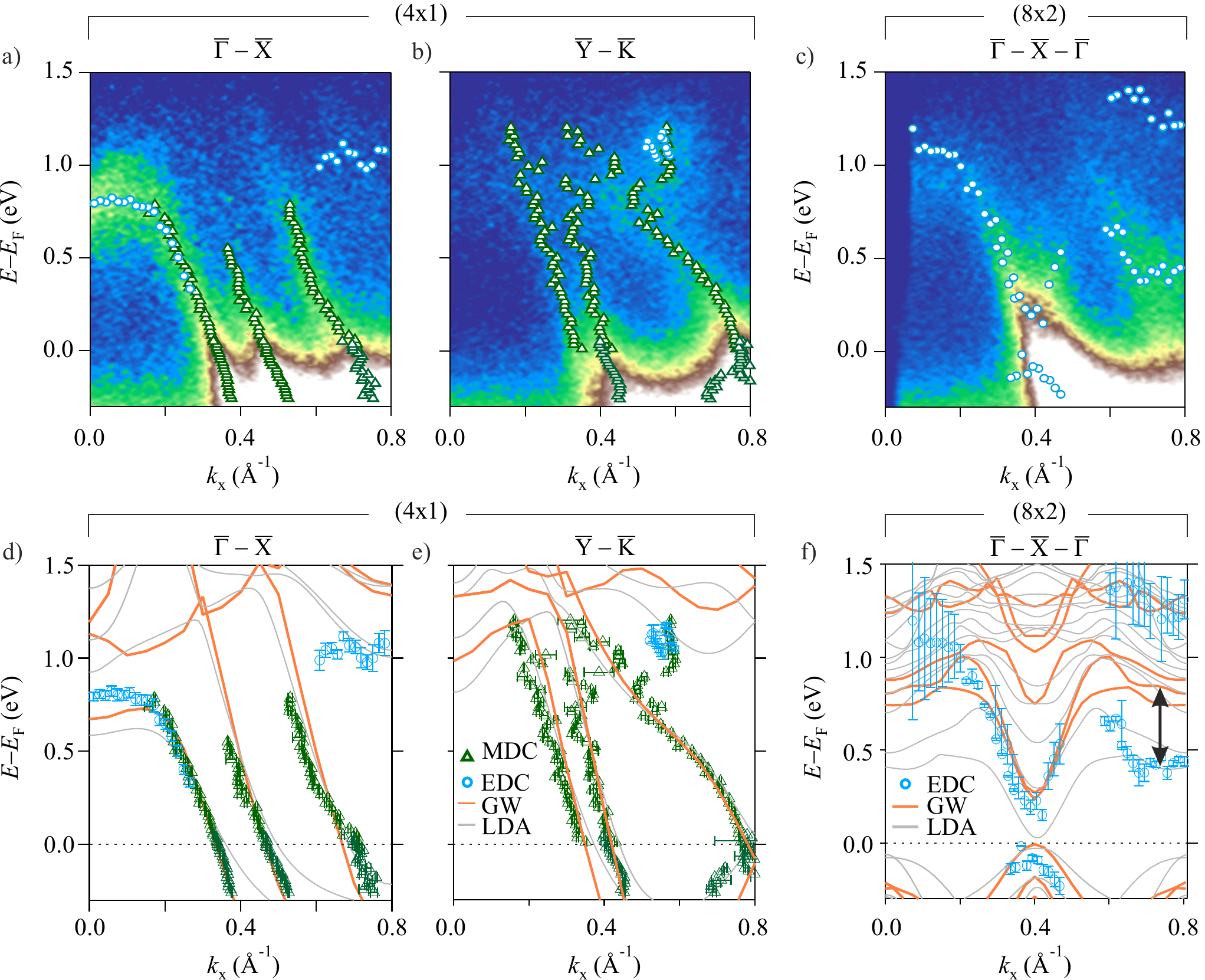}
\caption{\label{fig:DFT} a) - c) Electronic dispersions measured by trARPES along \Ga{} - \X{} and \Ga{} - \K{} in the (4x1) phase at 150~K and along \Ga{} - \X{} in the (8x2) phase at 25~K. Data are obtained at $\Delta t = 50$~fs with a fluence of 2~mJ cm$^{-2}$. Band positions extracted from fits are overlaid. d) - f) Extracted positions from a) - c) of the In/Si(111) band structure compared with DFT calculations; in the (4x1) phase along d) \Ga{} - \X{} and e) along \Ga{} - \K{} and f) in the (8x2) phase along \Ga - \X{}. The positions of the EDC fits reflecting the maximum of the fitted peak are marked in blue, while those obtained from MDCs are marked in green. The band positions are overlaid on the DFT calculated band structure within the LDA (gray) and GW (orange) approximations. The arrow in f) highlights the region where GW fails to capture the excited state positions.
}
\end{figure*}

We first investigate the unoccupied band structure of the In/Si(111) nanowires. Using trARPES at a fixed delay between pump and probe allows the normally unoccupied states up to the pump energy to be mapped, as shown in Fig.~\ref{fig:mapping} for In/Si(111) in the (4x1) phase. The data are again symmetrized about the $k_x$=0 line in order to show the bands within the entire Brillouin zone. While the trARPES data is measured in an area larger than the irreducible part of the Brillouin zone, such a symmetrized representation implicitly assumes constant photoemsission matrix elements on both sides of the Brillouin zone, which is likely not fulfilled. However, the position of bands, which is the quantity of interest at this point, will not be affected by this assumption. The constant energy slices are presented on a logarithmic color scale representing the photoemission intensity in order to capture the large dynamic range of intensity between occupied and excited states. These $k$-space distributions can be traced upwards in energy, gradually becoming less one-dimensional at higher energies, possibly due to the influence of coupling to, or overlap with, Si bulk states.

In order to analyze the details of the excited state dispersions, obtained at fixed $\Delta t$=50~fs, the ($E, k_{x}$) data are fitted to extract the band positions, the results of which are summarized in Fig.~\ref{fig:DFT}. A delay of 50~fs was chosen to be close to excitation while allowing some time for scattering and relaxation of electrons into lower energy states, in order to map as much of the unoccupied band structure as possible. An additional benefit is that it removes any influence of coherent effects produced during pump-probe overlap. Extracting the band positions allows a close comparison between the measured band structure and the predictions from DFT. Two types of distribution curves in momentum (MDC) and energy (EDC) are used to build a complete picture of the transiently occupied states. Fig.~\ref{fig:DFT}~a) and b) are obtained in the (4x1) phase along the two high-symmetry lines \Ga{}-\X{} and \Y{}-\K{} defined in Fig.~\ref{fig:thermalTrans}~a). The state dispersions can be clearly followed up to around 1~eV above \EF{} and are found to differ along the two high-symmetry directions in line with the predictions from theory. In the metallic (4x1) phase, the differences between LDA and GW excited states are small. Good overall agreement with the calculated band structure is obtained, with only minor differences: close to the \Ga{}-point, DFT predicts two bands at 0.65~eV and 1.1~eV. However, we cannot resolve these states due to the broad distributions in energy reflecting the intrinsically short lifetimes of these states. Additional states close to $k_{x}$~=~0.8~\AA$^{-1}$ appear substantially lower than the predicted GW bands along both high-symmetry directions; these may be Si states at the edge of the Si bulk band gap \cite{Gerstmann2014}.

In contrast, the comparison between data (Fig.~\ref{fig:DFT}~c)) and calculations in the (8x2) phase, shown in Fig~\ref{fig:DFT}~f), reveal a number of differences. A clear discrepancy between the LDA prediction and the data is the size of the \X{}-point band gap in the $m_{2}, m_{3}$ bands ($k_{x}$~=~0.38 \AA$^{-1}$): this is measured to be 270~meV in contrast to the LDA prediction of only 60~meV. Such deviations are a common feature in gapped systems described by LDA. In contrast, the GW calculation, which includes the dynamically screened Coulomb interaction between electrons beyond the mean-field approximation, produces a gap size of 240~meV, much closer to the experimental value. Indeed, the GW calculation well reproduces the majority of the band dispersion in the (8x2) phase, in stark contrast to the LDA results, as summarized in Table~\ref{tab:comp}. The fact that this gap and the majority of bands are so well reproduced by GW is remarkable. We note that the close agreement may result partially from a fortuitous cancellation of errors as discussed below \cite{Riefer2016}. However, significant deviations from the GW band structure are also observed. The states at \Ga{}$_{2,3}$ of the (8x2) Brillouin zone ($k_{x}$~=~0.8~\AA$^{-1}$) are found significantly lower than the GW bands (the fact that they fall on the LDA predicted band is most likely coincidental). This difference between the experimental band position and the GW calculation is considerable: around 350~meV. The discrepancy is even more striking as these states are expected to be symmetric to those at \Ga{} ($k_x$~=0) in the (8x2) unit cell, suggesting a $k$-dependent mechanism causing this band renormalization. Since these $m_{1}$ states are directly linked to the phase transition, in particular, as they shift down into the occupied region during the photo-induced phase transition \cite{Nicholson2018a}, it is tempting to think that the discrepancy may result from an incomplete initial transition into the (8x2) ground state which pulls the $m_{1}$ band down in energy. As described before, the measurements presented in Fig.~\ref{fig:thermalTrans}~c) do suggest that residual areas of surviving (4x1) phase persist even to lowest temperatures. However we can rule this out as the cause of the observed band position for two reasons. Firstly, the weak intensity of the residual $m_1$ band implies that the (8x2) phase clearly dominates at low-temperatures. Therefore the presence of small islands of the (4x1) phase cannot explain the band positions we find in the experiment, as such a mixed phase sample should lead to a superposition of the band structures of the two phases as measured by ARPES, but dominated by the (8x2) phase. Secondly, an incomplete transition of the (8x2) regions seems unlikely given that the gap at the \X{}-point has already fully developed and compares well with the GW prediction. 

We additionally rule out the variation of spectral weight of back folded bands in an extended zone scheme, discussed for the occupied states \cite{Gallus2001}, as the cause of the observed differences \cite{Nicholson2018a}. In the case of a CDW gap opening, spectral weight is expected to follow the bare dispersion \cite{Voit2000} hence some parts of the normalized dispersion appear only very weakly e.g. in Fig.~\ref{fig:thermalTrans}c) and Fig.~\ref{fig:SM_inputs}c). However this can only reduce the spectral weight of expected bands, and will not shift bands to energies not predicted by the back-folding, as we observe here. In order to motivate future investigations into these observations we discuss two main scenarios that may lead to such an effect: strain at the surface, and electron-hole attraction due to exciton formation.

\begin{table}[]
\caption{Comparison of excited state band positions in eV obtained from ARPES and DFT along \Ga{}--\X{}--\Ga{} in the (8x2) phase. DFT values are taken from the lowest of the predicted bands above \EF{}.}
\label{tab:comp}
\begin{tabular*}{\columnwidth}{@{\extracolsep{\fill}}ccccc}
\hline \hline
     $k_x$~(\AA$^{-1}$) & Expt. (eV) & LDA (eV) & GW (eV) \\ \hline
0.2  & 1.0 $\pm$ 0.1       & 0.46    & 0.82       \\
0.4  & 0.19 $\pm$ 0.05       & 0.03    & 0.24       \\
0.6  & 0.63 $\pm$ 0.05      & 0.46    & 0.82       \\
0.8  & 0.45 $\pm$ 0.05     & 0.40    & 0.74   \\ \hline \hline
\end{tabular*}
\end{table}

Strain has been shown to induce band structure changes on the order of hundreds of meV in transition metal dichalcogenides \cite{Lu2012, Castellanos-Gomez2013, Ghorbani-Asl2013, Amin2014}, although this can requires large strain values that may be difficult to achieve intrinsically in a material without externally applied force. Furthermore, strain has recently been shown capable of driving an insulator-to-metal transition in a correlated electron system \cite{Ricco2018a}. In the case of In/Si(111) it is well known that the displacement of particular atoms can have significant effects on the band structure \cite{Gonzalez2005, Wippermann2010, Jeckelmann2016} which could be partially induced by strain. In addition, strain induced by oxygen defects has been shown to modify the CDW transition \cite{Yeom2015} while strain at step edges influences the transport of electrons \cite{Hatta2017}. However calculations assuming a strain value of 1~\% along the atomic chain direction (compressive or tensile) produce changes to the $m_1$ band position of only 100~meV, which does not account for the effect we observe and tends to rule out strain as the dominant effect. Thus, while strain may play a minor role, it appears that additional effects have to be considered. Another possibility is that topological defects (solitons) may induce additional states that contribute to the photoemission signal, as suggested by STM data \cite{Cheon2015}. However it is unclear how such states may be observed with significant intensities in photoemission. 

A more likely scenario is that excitonic effects may arise in In/Si(111); in particular as the (8x2) phase is a reduced dimensional surface-confined system which is semiconducting. It is therefore conceivable that screening of the Coulomb interaction between electrons and holes is reduced compared with a bulk 3D material, leading to exciton formation with enhanced binding energies as in the family of transition metal dicholcogenides where such effects are large \cite{Chernikov2014}. As a comparison, the typical binding energies of 2D excitons on Si surfaces \cite{Rohlfing1999, Weinelt2004} are of the order of hundreds of meV, an order of magnitude higher than in the bulk, and are therefore a possible explanation of our observations. In the present case, however, the measurements indicate $k$-dependent exciton binding energies. Such effects are not included in the GW approximation. In order to examine these effects we performed occupation-constrained DFT calculations \cite{Frigge2017, Nicholson2018a}. In these calculations, holes and electrons are confined in $k$-space to specific valence and conduction band segments in the Brillouin zone, respectively. The comparison of the total-energy differences between the ground-state and occupation-constrained DFT calculations with the quasiparticle transition energies gives access to the exciton binding energies at the respective $k$-points. We calculate excitonic effects at \Ga{}$_2$ that are about 220~meV larger than at the \X{} point band gap, which would agree rather well with the measured findings. These calculations further suggest that the good agreement between the measured excitation energies and the GW calculations in particular at the \X{} point may partially result from an error cancellation between the missing self-consistency in the present GW calculations -- which is expected to increase considerably the excitation energies, see, e.g. Ref. \cite{Riefer2016} -- and the neglect of excitonic effects, which will reduce the excitation energies. The fact that the constrained DFT calculations predict exciton binding energies at \Ga{}$_2$ that are -- in good agreement with the measured data -- considerably larger than at the \X{} point band gap calls for further systematic investigations into the role that excitons play in the In/Si(111) system, in particular addressing the possibility of $k$-resolved excitonic coupling.

\subsection{2. Momentum-resolved population dynamics}

\begin{figure*}[ht]
\includegraphics[width=2\columnwidth]{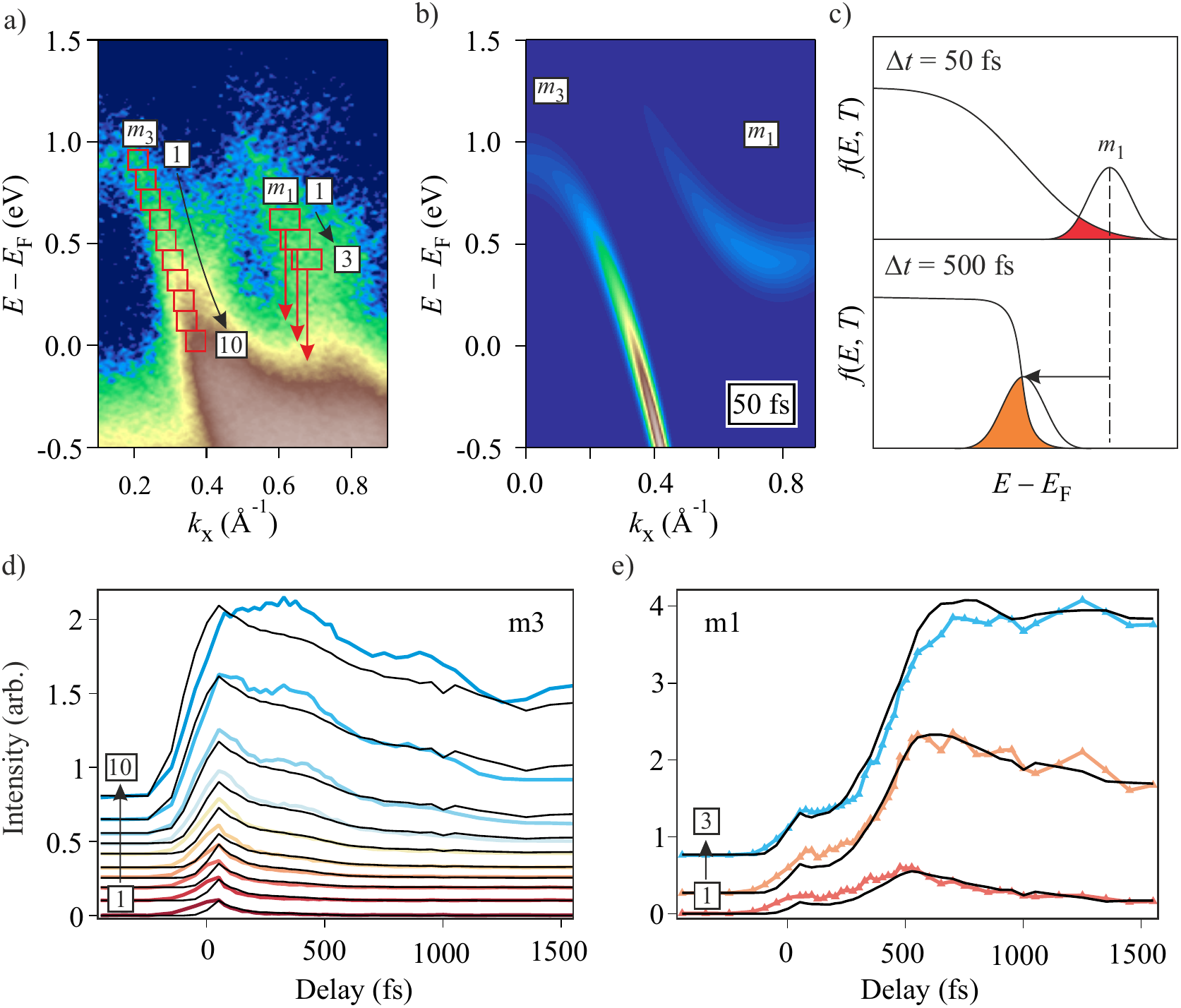}
\caption{\label{fig:lifetimes} a) Photoemission signal at a pump-probe delay of 50~fs. The ($E$, $k$) regions from which the population dynamics are extracted are marked as red boxes, which follow the dispersion of the $m_3$ and $m_1$ bands. The resulting intensity curves are shown in d) and e) and are compared with the thermal model of the spectral function, a single snap-shot of which is shown in b) at the electronic temperature corresponding to a delay of 50~fs. The solid black lines are the result of the thermal model for the corresponding region in the simulation. c) A schematic of the $m_1$ dynamics showing how the shift of the band leads to an increased thermal occupation over time. 
}
\end{figure*}

We turn now to the population dynamics of excited states and an analysis of the transient population dynamics within the In/Si(111)-(8x2) band structure. The evolution of the band structure itself during the PIPT has been addressed in our previous study \cite{Nicholson2018a}; here we focus on the $k$-resolved dynamics of the photoemission signal, and what this reveals about the relaxation of the excited carriers within the nanowires following photoexcitation. In particular, we will demonstrate that the momentum-resolved dynamics can be modeled by a thermal population of the electronic band structure, which reveals a highly elevated electronic temperature even after 1~ps in clear contrast to results from UED measurements which are sensitive to the dynamics and thermal excitation of the low energy lattice phonons. 

The motivation for carrying out a $k$-resolved thermal model for the electron population is to gain microscopic insight into the carrier dynamics and the flow of energy within the system using an intuitive approach. A completely general approach would be to employ collision integrals based on the Boltzmann equation \cite{Ashcroft1976} which describes all $k$-dependent microscopic scattering processes as well as electronic transport. However such a model would include a large number of unknown fitting parameters. Therefore a suitable simplification is to employ an ensemble description governed by a transient electron distribution function, which may initially be non-thermal. Such a population analysis based on distribution functions has the advantage that the ensemble nature of the description naturally reduces the complexity of the many possible scattering channels to a single meaningful descriptor; after electron thermalization the distribution function is governed by the Fermi-Dirac distribution with an electronic temperature. Microscopic insights are still possible by comparing the ensemble behavior of electrons with those of other excited sub-systems e.g. phonons, spins etc. This has a significant advantage over the standard approach of analyzing exponential time-constants of population decays as such a single time-constant may be the result of multiple cascading scattering processes which makes their physical interpretation in general difficult \cite{Yang2015}. We therefore employ a simple but microscopically insightful approach based on the transient electronic temperature of the system. While many trARPES studies have extracted a transient electronic temperature in a variety of materials, there are relatively few comparisons between the extracted electronic temperature and the $k$-resolved response of the system, \cite{Nicholson2016, Parham2017, Tengdin2018}. In principle such a comparison allows the validity and role of a quasi-equilibrium temperature to be assessed on ultrafast time scales. By comparing with other ultrafast techniques such as UED it is therefore possible to piece together the microscopic flow of energy between different degrees of freedom.

To investigate the population dynamics, the sample was excited with a fluence of 2~mJ cm$^{-2}$ and a time-delay series obtained. We note that since the pump does not excite direct transitions in the Si substrate the excitation must be primarily localized to the In layer. Population dynamics are obtained from the regions shown in Fig.~\ref{fig:lifetimes}~a) which follow the dispersions of bands $m_3$ and $m_1$; the resulting intensity traces are displayed in Fig.~\ref{fig:lifetimes}~d) and e). As detailed in our previous study \cite{Nicholson2018a}, the $m_1$ band shifts in energy during the PIPT. Thus, in order to address the intrinsic intensity changes at different energies within the band, the analyzed regions of the $m_1$ band are transiently shifted by the same amount. This shift is indicated by vertical arrows in Fig.~\ref{fig:lifetimes}~a) and the extracted band position used to define the simulation is shown in the Appendix (Fig.~\ref{fig:SM_inputs}~b)). As a result, the dynamics in the $m_1$ band originate from a combination of scattering processes towards lower energy within the band combined with the band shift induced by the PIPT, which results at the same time in an \textit{increase} in thermal occupation of the entire band, as shown schematically in Fig.~\ref{fig:lifetimes}~c). This leads to the large increase of the photoemission intensity around 550~fs after the initial excitation at all $k$-points in the band. Note that this maximum appears at the same delay at all energies, confirming a rigid shift of the band in this momentum range. The assignment of this behavior to the change in thermal occupation is supported by our model, as described in detail below. In contrast, the dynamics within the rigid $m_3$ band originate purely from the thermal relaxation(cooling) of the electrons. At energies far above the Fermi level, population is rapidly transferred to lower energy states where it gradually accumulates; in particular close to the Fermi level the excited electronic population is long-lived.

\begin{figure}[b]
\includegraphics[width=\columnwidth]{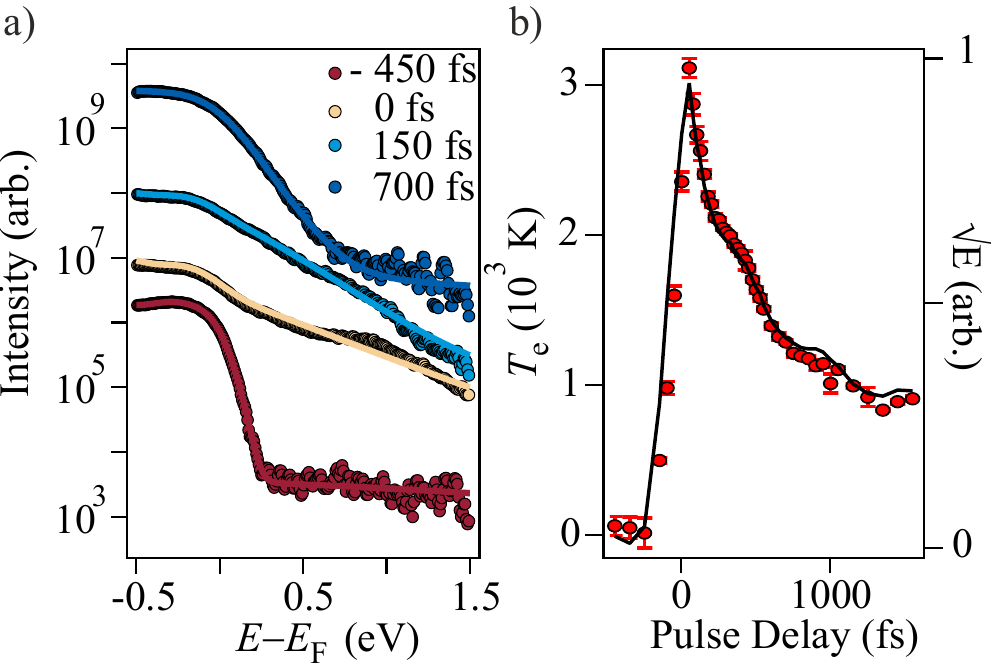}
\caption{\label{fig:elecTemp} a) EDCs in the region of the $m_1$ and $m_2$ bands and corresponding Fermi-Dirac fits at selected pump-probe delays. Only on very short time-scales (during the pump pulse) are significant deviations from thermal behavior observed. Curves are offset vertically for clarity. b) The transient electronic temperature curve extracted from the Fermi-Dirac function fits (data points). The extracted value of the electronic temperature is used to define the thermal population in the band structure simulations. The black curve is the square root of the excess energy in the unoccupied states as a function of delay, which is seen to closely follow the electronic temperature confirming a quasi-thermal distrubution. 
}
\end{figure}

In order to produce a thermal model of the electron population dynamics, we extract the electronic temperature from a Fermi-Dirac fit of an EDC integrated over a region of $\pm$0.15 \AA$^{-1}$ around the $m_3$ band at $k_x$~=~0.35~\AA$^{-1}$. Each EDC is fitted with a Gaussian peak with a constant background multiplied by a Fermi-Dirac function and convolved with the energy resolution. The peak is located at $-$0.28~eV binding energy and corresponds to the $m_2$/$m_3$ band which are merged together in the (8x2) phase. A small shift of maximum 0.09~eV is observed during photoexcitation; the width is assumed to be constant. The background obtained from the EDCs before excitation is held constant during the subsequent fitting at later delays. Representative fits and the resulting electronic temperature are shown in Fig.~\ref{fig:elecTemp}~a) and b). To confirm the validity of a quasi-equilibrium temperature we additionally plot the square root of the excess energy \cite{Nicholson2018a}, which is proportional to the electronic temperature of a thermalized system. The close agreement between the two quantities implies a quasi-thermal distribution is applicable in this case. In order to compare the measured population dynamics with the expectations of the thermal model, we define a model band structure consisting of two parabolic bands with band top (bottom) and effective mass corresponding to the $m_3$ ($m_1$) band as determined from our measurements of the excited state band structure presented in the subsection on excited state mapping. A Fermi-Dirac function is employed to define the electronic occupation at equilibrium at the experimental base temperature of 20~K. Ultrafast time dependence is included in the simulation by applying the Fermi-Dirac distribution for the transient electronic temperature corresponding to each time delay in the experiment to the model band structure, such that states above the Fermi level are populated following excitation. Therefore the spectral function is computed as $A(E, k,~\Delta t) = \epsilon(k,~\Delta t)$ x $f(E,~\Delta t)$, where $\epsilon(k,~\Delta t)$ is the dynamic band dispersion and $f(E,~\Delta t)$ is the Fermi-Dirac distribution at delay $\Delta t$. The resulting model spectral function is convolved by the experimental energy resolution of 150~meV and the result is shown in Fig.~\ref{fig:lifetimes}~b) at a time delay of 50~fs. 

Model population dynamics are extracted from equivalent regions and compared with the data in Fig.~\ref{fig:lifetimes}~d) and e) for the $m_3$ and $m_1$ bands respectively. In the case of the $m_3$ band, the transients show an initial rapid increase due to the pump excitation followed by gradual decline due to depopulation. The overall behavior is very well described by our model, particularly in the intermediate energy range where the scattering of the population to lower energies gradually reduces the electronic temperature. Certain features at both high and low energies do not match the thermal model, but the deviations are minor and the model captures the overall behavior well. At higher energies (boxes 1--5) and short times, there is a clear discrepancy due to the non-thermal population induced by the pump-pulse. However shortly after the end of the pump pulse the thermal model well describes the transient reduction of intensity. At low energies (boxes 7--10) a coherent oscillation becomes apparent in the data, resulting in a mismatch between the data and the model. Indeed we find clear signatures of the presence of coherent optical phonon modes in our data at different points in the band structure. A mode at 1.8~THz has been previously been observed in trARPES \cite{Chavez-Cervantes2018a}. The observation of coherent modes is clear evidence of the efficient excitation of optical phonon modes. However, we reiterate that overall the dynamics within this band appear to be well described by the evolution of the electron temperature and the thermal cooling of electrons.

The dynamics in the $m_1$ band show very different behavior. As discussed above, during the (8x2) to (4x1) transition in In/Si(111), the $m_1$ band shifts from above to below \EF{}, which should lead to an increase of the thermal occupancy of the band over time as it shifts towards \EF{}. Such an increase of intensity due to the lowering of the $m_1$ band in energy is clearly evident in Fig.~\ref{fig:lifetimes}~e) as a second maximum occurring at 550~fs. The model including the $m_1$ band shift (shown in the Appendix) reproduces the data including the second maximum extremely well, allowing us to clearly identify these dynamics as originating from the shift and increasing thermal population of this band, which occurs simultaneously with the cooling of the electrons.

The success of the simple thermal model in capturing the overall behavior of the $k$-resolved dynamics indicates the appropriateness of a quasi-equilibrium approach for the transient electronic population even to a strongly non-equilibrium scenario. In addition, it implies a very high energy content must remain in the electronic system after 1~ps, as indicated by the persisting elevated electron temperature of around 1000~K. We note that independent trARPES measurements find a similarly high and persistent electronic temperature on the picosecond time scale \cite{Chavez-Cervantes2018a}. Such a high energy content remaining in the electronic system seems to contradict the analysis of UED measurements for the same fluence range (1 -- 3 mJ cm$^{-2}$). In this work, a maximum increase of the lattice temperature of no more than 40~K was extracted from a Debye-Waller analysis of diffraction spots \cite{Frigge2018}. Furthermore, the lattice temperature was found to reach its maximum after 6~ps, which is considerably longer than the reduction in electronic temperature within 1~ps that we observe. While it is clear that the sensitivity of trARPES and UED to electrons and the lattice respectively may result in different measurable dynamics, such a significant difference in timescales strongly suggests that there is an third channel invisible to both trARPES and UED into which the electronic energy is transferred following excitation. The elastic scattering signal, which is utilized in a Debye-Waller analysis, depends on the atomic mean square displacement. As high-energy (optical) phonons have a smaller vibrational amplitude compared to low-energy modes, a time-dependent Debye-Waller analysis is particular sensitive to acoustic phonons and is thus thus prone to underestimate the population of high-energy phonons and consequently the rate of energy transfer from electrons to the lattice \cite{Waldecker2016, Waldecker2017}. It thus seems highly probable that the electronic energy is initially transferred preferentially into high-energy optical phonon modes, which are strongly coupled to the electronic system and act as a bottleneck for the cooling of the electrons. In this scenario, the transfer from hot optical phonons to lower-energy accoustic phonons occurs on longer time scales by phonon-phonon scattering. Given that a thermal model well reproduces the dynamics of the electronic system as shown above, this is strong evidence that the electronic system is in quasi-equilibrium with itself, and implies a highly non-thermal distribution within the phonon system on these time scales. Presumably the hot optical phonons remain out-of-equilibrium with the acoustic modes until around 6~ps, where UED finds the maximum lattice temperature to be reached. In fact, the importance of specific optical phonon modes for the PIPT in In/Si(111) has been pointed out previously \cite{Wippermann2010, Frigge2017, Lucke2018} and our finding underlines the importance of the combination of electron-phonon and phonon-phonon coupling in determining the energy flow dynamics within the In nanowires. 

A more detailed investigation of these dynamics will require more precise measurements between the regimes accessible with trARPES (electrons) and UED (atomic mean-square displacement) in order to determine the non-equilibrium distribution of the optical phonons. One possibility is to employ time-resolved inelastic electron scattering, which provides momentum-resolved information on phonon dynamics \cite{Weathersby2015, Waldecker2017b, Stern2018a}. For example, a recent study employed MeV UED to follow the energy transport between electronic to phononic systems by comparing the temporal evolution of reflection peaks and the thermal diffuse background, allowing the dynamic population of specific sub-sets of phonons to be experimentally determined \cite{Konstantinova2018}. In the case of \InSi{} we observe clear evidence for a non-thermal phonon population that could in principle be quantified by combining the results of trARPES with ultrafast elastic and inelastic electron scattering. By using the input from both experimental probes, it should be possible to quantify and access the microscopic couplings between specific modes that result in the observed dynamics, which would go beyond an ``N-temperature'' description, where the potentially complex interactions are reduced to a single coupling constant. This also calls for more detailed calculations of the dynamic electron-phonon and phonon-phonon coupling in In/Si(111) in order to determine which modes are most relevant for ultrafast energy transfer in this model reduced-dimensional system.

\section{IV. Summary}

By combining time, energy and momentum resolution via trARPES to investigate excited states and ultrafast dynamics in In/Si(111) nanowires, we have gained detailed insights into both the electronic structure and the dynamic evolution of electronic population. In particular, by comparing with the calculated band structure in both (4x1) and (8x2) phases we have found evidence for additional effects not accounted for at the level of DFT. This may be attributed to the signature of a $k$-dependent interaction -- presumably excitonic in nature -- which may be enhanced in this low-dimensional system. In addition, by extracting momentum-resolved transient populations and comparing to a spectral function populated by a thermal electron distribution, we disentangle the effects of electronic cooling and rigid band motion during the PIPT in this system. This highlights the appropriateness of a transient thermal ensemble description during the PIPT, and by comparison with UED measurements suggests the existence of a non-equilibrium phonon distribution dominated by high-energy optical phonons. We anticipate that future work will allow the quantification and microscopic understanding of both the band structure and the microscopic couplings that govern the energy flow between the various sub-systems in this model system.

Our results provide significant new insights into an important model phase transition at a surface, and highlight trARPES as a powerful multi-dimensional spectroscopy offering complementary insights to a range of both equilibrium ARPES and ultrafast probes.

\section{Acknowledgments}

We acknowledge S. Wippermann for fruitful discussions. We gratefully acknowledge funding from the Max- Planck-Gesellschaft, and the Deutsche Forschungsgemeinschaft through FOR1700 and TRR142. The Paderborn Center for Parallel Computing (PC$^2$) and the H\"ochstleistungs-Rechenzentrum Stuttgart (HLRS) are acknowledged for grants of high-performance computer time. C.W.N. acknowledges support by the Fonds National Suisse pour la Recherche Scientifique through Div. II.

\section{Appendix}

\begin{figure}
\includegraphics[width=\columnwidth]{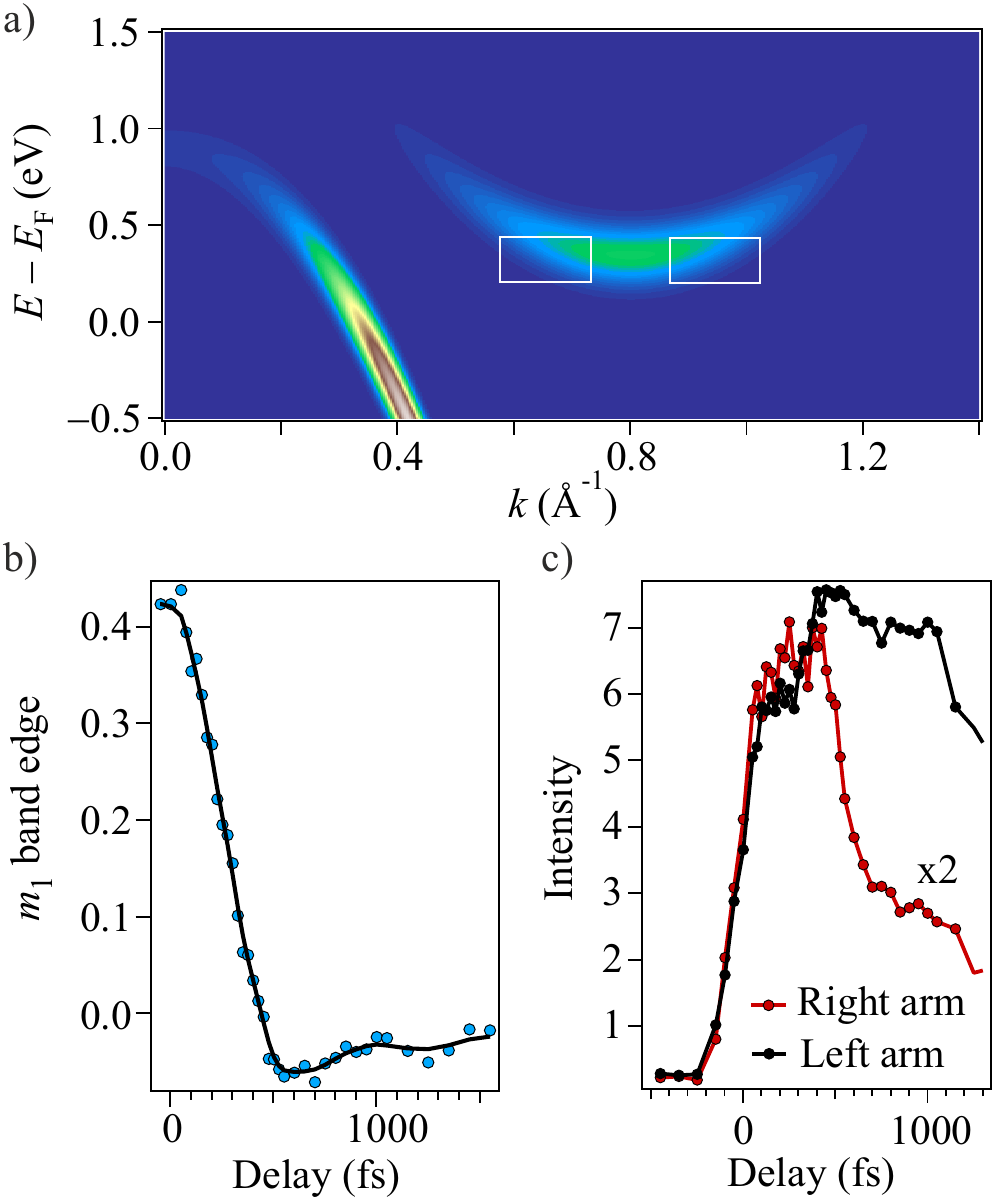}
\caption{\label{fig:SM_inputs} a) Simulated spectral function at a delay of 50~fs. Boxes mark the regions of the $m_1$ band that are analyzed in c). b) Extracted shift of the $m_1$ band as a function of delay. This shift is used as an input to the simulation and results in the increased thermal population of the band discussed in the main text. c) Population transients centered at (0.2~eV, 0.7~\AA$^{-1}$) and (0.2~eV, 0.8~\AA$^{-1}$) in the $m_1$ band of the (8x2) phase. The asymmetry after 400~fs suggests the removal of spectral weight from the right arm due to the PIPT.
}
\end{figure}

\textit{Rigid band shift --} In addition to the quasi-equilibrium electronic temperature, the rigid shift of the $m_1$ band over time is included in the model in order to reproduce the effects of the PIPT (see Fig.~\ref{fig:SM_inputs}~b)). The band position is determined from a Gaussian fit of the EDC obtained at the band bottom at $k_x~=~0.75$~\AA$^{-1}$. A shift of 0.85~meV/fs is observed. This shift leads to the increase in population observed in Fig.~\ref{fig:lifetimes}~e) as described in the main text.

\textit{Redistribution of spectral weight --} Changes to the distribution of spectral weight within the spectral function can also influence the population dynamics within a band. Such an effect caused by the PIPT can be seen by comparing the population in the two arms of the parabolic $m_1$ dispersion in the (8x2) phase as a function of delay, shown in Fig.~\ref{fig:SM_inputs}~c). The traces are extracted from static regions centered on (0.2~eV, 0.7~\AA$^{-1}$) and (0.2~eV, 0.8~\AA$^{-1}$) as shown schematically in a). After excitation both bands show a plateau behavior until 400~fs, as a result of state filling from higher lying bands balancing the removal of population. After 400~fs there is a sudden deviation as the population in the right arm decreases sharply while the left arm decreases only gradually. This can be explained by a reduction of the spectral weight in the right arm due to the change in the underlying symmetry during the completion of the phase transition. However, such considerations do not alter the conclusions of the electronic temperature analysis presented in the main text.

%

\end{document}